\documentclass[aps,twocolumn,floatfix,amsmath,amssymb,superscriptaddress]{revtex4-2}
\usepackage{amsmath}
\usepackage{graphicx}
\usepackage{siunitx}
\usepackage{braket}
\usepackage{booktabs} %for table
\usepackage{multirow} % to allow merged rows
\usepackage[usenames,dvipsnames]{xcolor}

\usepackage[caption = false]{subfig}

\usepackage{hyperref}
\hypersetup{colorlinks=true, urlcolor=blue, linkcolor=blue, citecolor=blue}

\usepackage{color}
\usepackage{float}
\usepackage{placeins}
\definecolor{purple}{rgb}{0.5,0,0.5}

\begin{document}

\title{First Principles Assessment of CdTe as a Tunnel Barrier at the $\mathbf{\alpha}$-Sn/InSb Interface}
%or Investigation of InSb/CdTe/Sn interfaces via density functional theory

% mjj45@pitt.edu
\author{Malcolm J. A. Jardine}
\thanks{These authors contributed equally to this work}
\affiliation{Department of Physics and Astronomy, University of Pittsburgh, Pittsburgh, PA, 15260, USA}

\author{Derek Dardzinski}
\thanks{These authors contributed equally to this work}
\affiliation{Department of Materials Science and Engineering, Carnegie Mellon University, Pittsburgh, PA 15213, USA}

\author{Maituo Yu}
\affiliation{Department of Materials Science and Engineering, Carnegie Mellon University, Pittsburgh, PA 15213, USA}

\author{Amrita Purkayastha}
\affiliation{Department of Physics and Astronomy, University of Pittsburgh, Pittsburgh, PA, 15260, USA} 

\author{A.-H. Chen}
\affiliation{Univ. Grenoble Alpes, CNRS, Grenoble INP, Institut N\'eel, 38000 Grenoble, France} 

\author{Yu-Hao Chang}
\affiliation{Materials Department, University of California-Santa Barbara, Santa Barbara, CA, USA}

\author{Aaron Engel}
\affiliation{Materials Department, University of California-Santa Barbara, Santa Barbara, CA, USA}

\author{Vladimir N. Strocov}
\affiliation{Paul Scherrer Institut, Swiss Light Source, CH-5232 Villigen PSI, Switzerland}

\author{Moïra Hocevar}
\affiliation{Univ. Grenoble Alpes, CNRS, Grenoble INP, Institut N\'eel, 38000 Grenoble, France}

\author{Chris J. Palmstr{\o}m}
\affiliation{Materials Department, University of California-Santa Barbara, Santa Barbara, CA, USA}
\affiliation{Department of Electrical and Computer Engineering, University of California-Santa Barbara, Santa Barbara, CA, USA}

\author{Sergey M. Frolov}
\affiliation{Department of Physics and Astronomy, University of Pittsburgh, Pittsburgh, PA, 15260, USA} 

\author{Noa Marom}
\email{Corresponding author: nmarom@andrew.cmu.edu}
\affiliation{Department of Materials Science and Engineering, Carnegie Mellon University, Pittsburgh, PA 15213, USA}
\affiliation{Department of Physics, Carnegie Mellon University, Pittsburgh, PA 15213, USA}
\affiliation{Department of Chemistry, Carnegie Mellon University, Pittsburgh, PA 15213, USA}

\begin{abstract}
Majorana zero modes, with prospective applications in topological quantum computing, are expected to arise in superconductor/semiconductor interfaces, such as $\beta$-Sn and InSb. However, proximity to the superconductor may also adversely affect the semiconductor's local properties. A tunnel barrier inserted at the interface could  resolve this issue. We assess the wide band gap semiconductor, CdTe, as a candidate material to mediate the coupling at the lattice-matched interface between $\alpha$-Sn and InSb. To this end, we use density functional theory (DFT) with Hubbard U corrections, whose values are machine-learned via Bayesian optimization (BO) [npj Computational Materials 6, 180 (2020)]. The results of DFT+U(BO) are validated against angle resolved photoemission spectroscopy (ARPES) experiments for $\alpha$-Sn and CdTe. For CdTe, the z-unfolding method [Advanced Quantum Technologies, 5, 2100033 (2022)] is used to resolve the contributions of different $k_z$ values to the ARPES. We then study the band offsets and the penetration depth of metal-induced gap states (MIGS) in bilayer interfaces of InSb/$\alpha$-Sn, InSb/CdTe, and CdTe/$\alpha$-Sn, as well as in tri-layer interfaces of InSb/CdTe/$\alpha$-Sn with increasing thickness of CdTe. We find  that 16 atomic layers (3.5 nm) of CdTe can serve as a tunnel barrier, effectively shielding the InSb from MIGS from the $\alpha$-Sn. This may guide the choice of dimensions of the CdTe barrier to mediate the coupling in semiconductor-superconductor devices in future Majorana zero modes experiments.

\end{abstract}

\maketitle

\section{Introduction}

A promising route toward the realization of fault-tolerant quantum computing schemes is developing materials systems that can host topologically protected Majorana zero modes (MZMs)~\cite{2016_AAsen_MilestonesMajoranhysQuantumComputing_PhysRevX.pdf,Sarma_MZM_topological_QC}. MZMs may appear in one-dimensional topological superconductors \cite{2010_Oreg_HelicalLiquidMBS_PRL,2010_Lutchyn_Semi-SuperHybridModelIntialPaper_PRL,2012_Mourik_SignaturesMZMHyridDevices_Sci}, which can be effectively realized by proximity coupling a conventional superconductor and a semiconductor nanowire that possesses strong spin-orbit coupling (SOC). Adding in a magnetic field enables this system to behave as an effective spinless p-wave topological superconductor, which allows for MZM states \cite{2012_Leijnse_IntroTopoSCandMF_IOP}. Recently, there have been new developments in material choices and experimental methods to identify MZMs in semiconductor nanowire-superconductor systems \cite{2021_Fu_ExperimentalReviewMZMHybridNanowires_SciChinaPhys}, designed to overcome challenges identified during the first wave of experiments 
%such as disorder and excessive coupling between the metal and semiconductor
\cite{YuNatPhys2021,2012_Kells_Near0EnergyStatesNotMZMSmoothConfinement_PRB,2020_Frolov_ReviewTopoSCHybridDevices_NatPhys}. These include trying new combinations of semiconductors and epitaxial superconductors, e.g. Pb, Sn, Nb, to maximize the electron mobility and utilize larger superconducting gaps and higher critical magnetic fields \cite{2020_Khan_SCShadowJunctionInAsInSbxAlSnPn_ACSNano,2021_Pendharkar_MagFieldResistentInSbSnShells_ScienceReports,2015_Krogstrup_Epitaxy_SM_SCNanowires_NatMaterials,2020_Veld_GrowInAsonAlNanowire_CommPhys,2015_Krogstrup_Epitaxy_SM_SCNanowires_NatMaterials,2020_Carrad_ShadowEpiytaxySCSMInterfaces_AdvMat,2021_Kanne_EpitaxialPbOnInAsNanowires_NatNano}. Additionally, new proposed architectures include creating nanowire networks and inducing the field via micromagnets \cite{YJunction_Braidinng_PhysRevResearch,2021_Jardine_MicromagneticMuMaxMZMModel_SciPost}.

One of the challenges presented by the superconductor/semiconductor nanowire construct, is that excessive coupling between the superconducting metal and semiconductor may “metallize” the semiconductor, thus rendering the topological phase out of reach. Theoretical studies that treated the semiconducting and superconducting properties via the Poisson-Schr\"{o}dinger equation, have shown that excessive coupling between the materials may lead to the semiconductor's requisite properties, such as the Lande\'e g-factor and spin-orbit-coupling (SOC), being renormalized to a value closer to the metal's. In addition, large unwanted band shifts may be induced \cite{2018_Reeg_MetallizationOfSMBySCInterface_PRB,2018_Antipov_EffectGateMajoranaNanowires_PRX,2021_Pendharkar_MagFieldResistentInSbSnShells_ScienceReports,2021_Vail_GatedMagnetotransportIaSnCdTe_Springer,2017_Reeg_SC-SMModelConectsToExperimentSOCEnergy_PRB}. Having a tunnel barrier could modulate the superconductor-semiconductor coupling strength and thus the induced proximity effect, which is critical for controlling experiments. It is currently unknown what the required width range of a tunnel barrier is. Another potential benefit of a CdTe layer is InSb surface passivation.

InSb and Sn are among the materials used to fabricate devices for Majorana search~\cite{2019_Badawy_HighMobiltyInSbNanowires_NanoLet}. InSb is the backbone of such systems because it has the highest electron mobility, strongest spin-orbit coupling (SOC) and a large Landé g-factor in the conduction band compared to other III-V semiconductors. $\beta-$Sn has a bulk critical field of 30 mT and a superconducting critical temperature of 3.7 K, higher than the 10 mT and 1 K, respectively, of Al. Recently, $\beta$-Sn shells have been grown on InSb nanowires, inducing a hard superconducting gap \cite{2021_Pendharkar_MagFieldResistentInSbSnShells_ScienceReports}. The large band gap semiconductor CdTe is a promising candidate to serve as a tunnel barrier. Thanks to its relative inertness, it may simultaneously act as a passivation layer protecting the InSb from environmental effects and potentially minimizing disorder \cite{2015_Cole_SoftHardGapEffectsSMMZM_PhysRevB,2022_Badawy_EpitaxyCdTeOnInSbNanowires_ADVSCI}. Advantageously, CdTe is lattice matched to InSb  \cite{2015_Negreira_DFTSnMonolayeronInSbCdTe}. Sn has two allotropes. The $\beta$ form, with a BCT crystal structure, is of direct relevance to MZM experiments thanks to its superconducting properties. However, the semi-metallic $\alpha$ form has a diamond structure, which is lattice matched to InSb and CdTe, making it an ideal model system for investigating,  both theoretically and experimentally, the electronic structure of Sn/InSb heterostructures. 

Much experimental work, such as growth and ARPES studies, has been undertaken on $\alpha$-Sn. Previously, $\alpha$-Sn has been found to possess a topologically trivial band inversion, with SOC inducing a second band inversion and a topological surface state (TSS) \cite{2017_Rogalev_DoubleBandInversionaSnOrbital_PRB,2022_InSbaSn001ARPESCalibrationTicknessDependent_PRB}. The effect of strain on the topological properties of $\alpha$-Sn has also been studied \cite{2020_Shi_FirstprincipleTopologicalSurfaceaSn111_Elsevier.pdf,2017_Xu_TopoDiracSemimetalaSnInSb111_PRL,2021_Vail_GatedMagnetotransportIaSnCdTe_Springer,2017_Huang_StrainedSnDiracSemimetalMagnetoresitance_PRB,2020_Vail_GrowaSnCdTe_PSS,2018_Coster_HamiltonianStudyaSnonCdTeThinFilms_PRB,2020_Madarevic_StructuralElectronicPropertiesaSn_APLMaterials,2013_Barfuss_TopoInsulatoraSnInSb001_PRL,2018_Scholz_TopoSurfaceStateaSnPhotoemission_PRB,2013_Ohtsubo_DiracConeHelicSpinExperimentThinaSn001InSb_PRL,2007_Fu_TopoInsulatorsInversionSymm_PRB}. In-plane compressive strain has been reported to make $\alpha$-Sn a topological Dirac-semi-metal and induce a second TSS to appear \cite{2017_Rogalev_DoubleBandInversionaSnOrbital_PRB}. Conversely, tensile strain has been reported to induce a transition to a topological insulator. CdTe \cite{2022_Badawy_EpitaxyCdTeOnInSbNanowires_ADVSCI} and $\alpha$-Sn \cite{2021_Pendharkar_MagFieldResistentInSbSnShells_ScienceReports,2022_InSbaSn001ARPESCalibrationTicknessDependent_PRB} have been epitaxially grown on InSb. Depositing Sn on InSb often leads to growth of epitaxially matched $\alpha$-Sn, although $\beta$-Sn may appear under some conditions \cite{2023_Khan_arXiv}. In addition, $\alpha$-Sn can transition to $\beta$-Sn if the Sn layer is above a critical thickness or if heat is applied during fabrication processes \cite{1981_Farrow_GrowingSnheteroepitaxial_JourCrysGrowth,2019_Huanhuan_GrowingSnInSbCdTe_AdvEnginMat}. Studying the interface with the lattice matched $\alpha$-Sn may provide insight, which is also pertinent to $\beta$-Sn as both could be present in hybrid systems. Therefore, these are promising materials to investigate for future device construction.

MZM experiments rely on finely tuned  proximity coupling between a superconducting metal and a semiconductor. By adding a tunnel barrier at the interface between the two materials and varying its width, one could potentially mediate the proximity coupling strength to achieve precise control over the interface transparency. To the best of our knowledge, this idea has not yet been tested in experiments and it is presently unknown which material(s) would be the best choice for a barrier and what would be the optimal thickness. Simulations of a tri-layer system with a tunnel barrier are therefore needed to inform MZM experiments.  Here, we use density functional theory (DFT) to study a tri-layer system, in which InSb is separated from $\alpha$-Sn by a  CdTe tunnel barrier. Despite recent progress towards treating superconductivity within the framework of DFT \cite{2020_Davydov_PRB, 2020_Sanna_PRL} the description of proximity-induced superconductivity at an interface with a semiconductor is still outside the reach of present-day methods. However, DFT can provide useful information on properties, such as the band alignment at the interface. Conduction band offsets are of particular importance because the proximity effect in most experiments on InSb primarily concerns the conduction band. In addition, DFT can provide information on the penetration depth of metal induced gap states (MIGS) into the semiconductor  \cite{2022_Badawy_EpitaxyCdTeOnInSbNanowires_ADVSCI,2019_Winkler_NumericalSMSCHetrostructureModel_PRB,2018_Antipov_EffectGateMajoranaNanowires_PRX,2021_Schuwalow_SMSCInterfaceBandOffsetInAsAl_AdvSci}, which is important for determining the appropriate thickness of the tunnel barrier. 

Within DFT, computationally efficient (semi-)local exchange-correlation functionals severely underestimate the band gap of semiconductors to the extent that some narrow-gap semiconductors, such as InSb, are erroneously predicted to be metallic \cite{2019_Borlido_ExchangeCorrelationFunctionalsBenchmarkBandGaps_JCHEMTHEORYCOMPUT,2019_Bennett_HubbardUGBRVUltrasoftPseudopotential_CompMatSci,2016_Huang_DFTUaFe0_JPHYSCHEMUS,2012_Peverati_M11-LFunctionalBandgapsSemiconductor_JCHEMPHYS}. This is attributed to the self-interaction error (SIE), a spurious repulsion of an electron from its own charge density \cite{1981_Perdew_SelfInteractionError_PhysRevA,2006_Sanchez_SelfInteractionError_JourChemPhys,2008_Sanchez_SelfInteractionErrorBandGaos_PhysRevLett}. Hybrid functionals, which include a fraction of exact (Fock) exchange, mitigate the SIE and yield band gaps in better agreement with experiment. However, their computational cost is too high for simulations of large interface systems, such as the $\alpha$-Sn/CdTe/InSb tri-layer system studied here. The DFT+U approach, whereby a Hubbard U correction is added to certain atomic orbitals, provides a good balance between accuracy and computational cost\cite{2019_Borlido_ExchangeCorrelationFunctionalsBenchmarkBandGaps_JCHEMTHEORYCOMPUT,2018_Zhang_DFTFunctionalCohesiveProperties_NewJPhys.pdf,2016_Garza_HybridFunctionalBenchmark_JPhyChemLet}. Recently, some of us have proposed a method of machine learning the U parameter for a given material by Bayesian optimization (BO) \cite{2020_YuMarom_BayesianOptimizationHubbardU_npjCompMaterials}. The DFT+U(BO) method has been employed successfully for InSb and CdTe  \cite{2021_Yang_FirtsPrincipleAssessInAs-GaSbInterfaceCdTe_PRM.pdf}. 

It has been shown that (semi-)local functionals fail to describe the bulk band structure of $\alpha$-Sn correctly, specifically the band ordering and the orbital composition of the valence bands at the $\Gamma$ point. DFT+U, hybrid functionals, or many-body perturbation theory within the GW approximation are necessary to obtain a correct description of the band structure \cite{1998_Rohlfing_dElectronsForBSCalsaSn_PRB,2013_Kufner_StructuralElectronicPropertiesaSnNanocrystals_PRB,2013_Barfuss_TopoInsulatoraSnInSb001_PRL,2020_Shi_FirstprincipleTopologicalSurfaceaSn111_Elsevier.pdf,2022_Dardzinski_BestPracticesDFTInorganicInterfacesBO_JPHYS-CONDENSMAT}. DFT+U simulations have required slab models of more than 30 monolayers of Sn to converge towards a bulk regime, where quantum confinement is no longer dominant. With a small number of layers $\alpha$-Sn may exhibit topological properties \cite{2018_Xu_StaneneInSb111ElectronicStructureUValue_PRB,2015_Negreira_DFTSnMonolayeronInSbCdTe,2014_Kufner_TopoaSnSurfaceStatesVsThicknessStrain_PRB}. Some DFT studies have considered slab models of bi-axially strained $\alpha$-Sn. DFT simulations of strained $\alpha$-Sn  on InSb have been conducted with a small number of layers of both materials \cite{2019_Rogalev_TailoringTopoSurfaceStateaSnFilms_PRB,2015_Negreira_DFTSnMonolayeronInSbCdTe}. The DFT+U approach has reproduced the effects of strain and compared well with experimental data \cite{2022_InSbaSn001ARPESCalibrationTicknessDependent_PRB,2018_Xu_StaneneInSb111ElectronicStructureUValue_PRB,2019_Rogalev_TailoringTopoSurfaceStateaSnFilms_PRB}. 

Here, we perform first principles calculations using DFT+U(BO) for a (110) tri-layer semiconductor/tunnel barrier/metal interface composed of the materials InSb/CdTe/$\alpha$-Sn, owing to their relevance to current Majorana search experiments \cite{2022_Badawy_EpitaxyCdTeOnInSbNanowires_ADVSCI,2021_Pendharkar_MagFieldResistentInSbSnShells_ScienceReports}. To date, DFT studies of large interface slab models with a vacuum region have not been conducted for these interfaces. Previously, the results of DFT+U(BO) for InSb(110) have been shown to be in good agreement with angle-resolved photoemission spectroscopy (ARPES) experiments \cite{2022_Yang_TransforamtionMatrixBandUnfoldingInAsInSbSurfaces_AdvQuanTech}. Here, we also compare the results of DFT+U(BO) to ARPES for $\alpha$-Sn (Section \ref{Sn}) and CdTe (Section \ref{CdTe}). Excellent agreement with experiment is obtained. In particular, for CdTe the z-unfolding scheme (Section \ref{Z-Unfolding}) helps resolve the contributions of different $k_z$ values and modelling the 2 $\times$ 2 surface reconstruction reproduces the spectral signatures of surface states. We then proceed to study the bi-layer interfaces of InSb/CdTe, CdTe/$\alpha$-Sn, and InSb/$\alpha$-Sn (Section \ref{bilayer}).  Finally, to  assess the effectiveness of the tunnel barrier, we study tri-layer interfaces with 2 to 16 monolayers (0.5 nm to 3.5 nm) of CdTe inserted between the InSb substrate and the $\alpha$-Sn (Section \ref{tri-layer}). This thickness is within the thickness range of CdTe shells grown on InSb nanowires. For all interfaces, our simulations provide information on the band alignment and the presence of MIGS. 
We find that 16 layers of CdTe (about 3.5 nm) form an effective tunnel barrier,  insulating the InSb from the $\alpha$-Sn. However, this may be detrimental for transport at the interface. Based on this, we estimate that the relevant thickness regime for tuning the coupling between InSb and Sn may be in the range of 6-10 layers of CdTe.

\section{METHODS} 

\subsection{Z-Unfolding}\label{Z-Unfolding}

\begin{figure*}[t]
\centering
\includegraphics[width=1\textwidth]{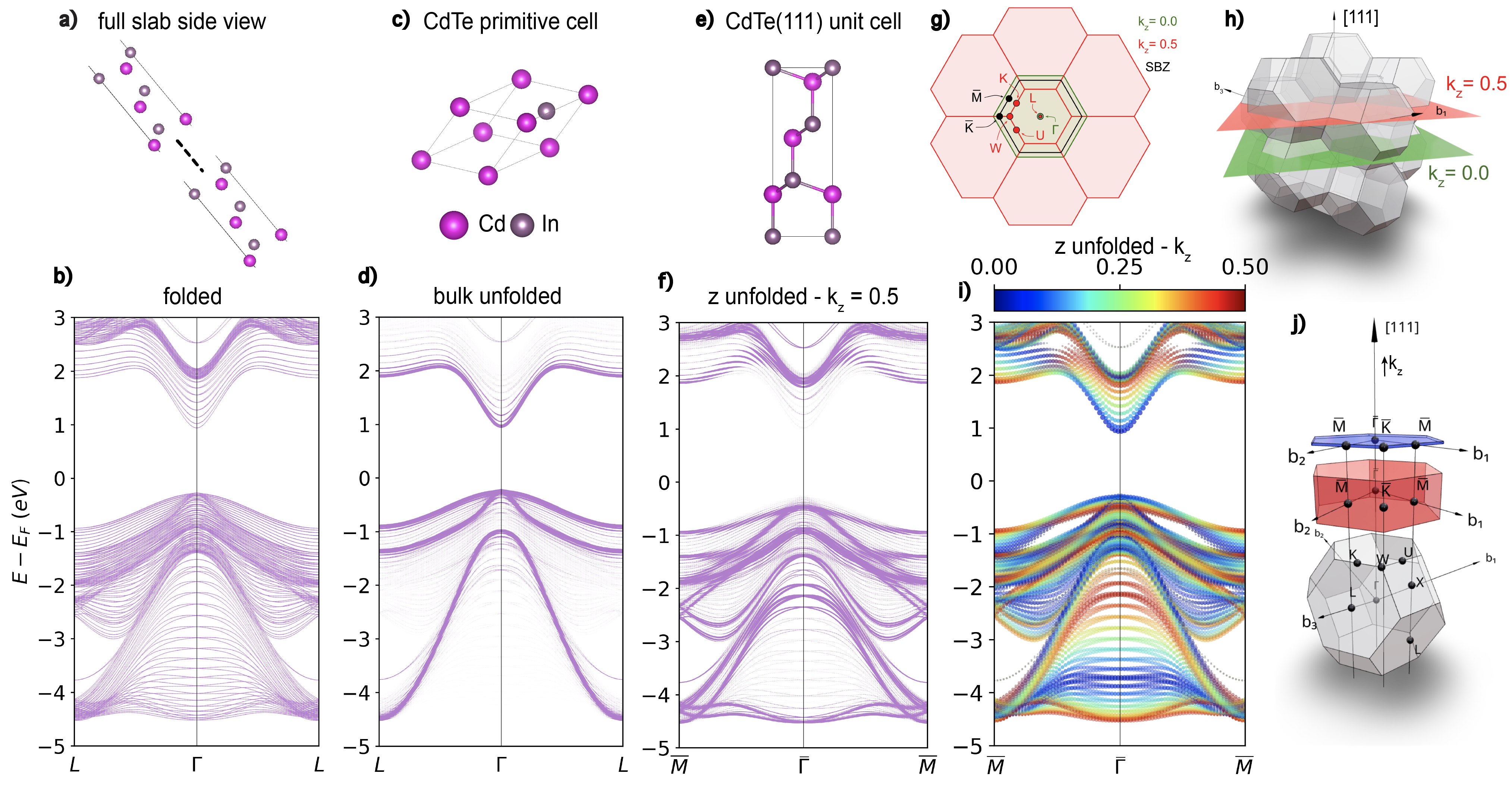}
\caption{
(a) Side view of the CdTe(111) slab (b) Folded band structure of CdTe(111) 25 monolayer slab. (c) Primitive unit cell of CdTe (d) bulk-unfolded band structure (e) unit cell of CdTe(111) slab used in z-unfolding.
(f) Z-unfolded band structure along k-path $\overline{M}-\overline{\Gamma}-\overline{M}$ for $k_z=0.5$, and (g) as a function of $k_z$.
(h) FCC bulk BZ (grey), (111) unit-cell BZ (red) and (111) surface BZ (blue). 
(i) Intersecting planes slice through the bulk BZ for $k_z=0$ (green) and $k_z=0.5$ (red) with the SBZ indicated.
(j) tessellated bulk BZs showing (111) orientated intersecting planes for given $k_z$ values. 
\label{Fig:z_unfolding_illustration}}
\end{figure*}

Simulations of large supercell models produce complex band structures with a large number of bands, as shown in Figure~\ref{Fig:z_unfolding_illustration}a,b for a CdTe(111) slab with 25 atomic layers, whose band structure was calculated using PBE+U(BO), as described in Section \ref{computational}. Band structure unfolding is a method of projecting the band structure of a supercell model onto the appropriate smaller cell (\cite{2012_Popescu_BandUnfoldingBackground_PRB,2010_Ku_BandUnfoldingBackground2_PRL,2020_Herath_PyProcarCode_ComputerPhysicsCommunications,2021_Wang_VASPKIT_ComputerPhysicsCommunications,2014_Medeiros_BandUnfoldingBackground3_PRB, 2022_Yang_TransforamtionMatrixBandUnfoldingInAsInSbSurfaces_AdvQuanTech}. This can help resolve the contributions of states emerging from of \textit{e.g.,} defects and surface reconstructions vs. the bulk bands of the material. In addition, it can facilitate the comparison to angle-resolved photoemission spectroscopy (ARPES) experiments. The ``bulk band unfolding" scheme \cite{2022_Yang_TransforamtionMatrixBandUnfoldingInAsInSbSurfaces_AdvQuanTech} projects the supercell band structure onto the primitive unit cell, illustrated in Figure \ref{Fig:z_unfolding_illustration}c. The resulting band structure, shown in Figure \ref{Fig:z_unfolding_illustration}d, appears bulk-like. Bulk-unfolded band structures have been shown to compare well with ARPES experiments using high photon energies, which are not surface sensitive owing to the large penetration depth.

The ``z-unfolding" scheme \cite{2022_Yang_TransforamtionMatrixBandUnfoldingInAsInSbSurfaces_AdvQuanTech} projects the band structure of a slab model with a finite thickness onto the Brillouin zone (BZ) of a single layer of the slab supercell with the same orientation, illustrated in Figure~\ref{Fig:z_unfolding_illustration}e. The resulting band structure, shown in Figure~\ref{Fig:z_unfolding_illustration}f, contains extra bands that are not present in the bulk-unfolded band structure. The extra bands originate from different $k_z$ values in the 3D primitive Brillouin zone projecting onto the surface Brillouin zone (SBZ), creating overlapping paths. For example, panel Figure~\ref{Fig:z_unfolding_illustration}g shows cross sections through the BZ at values of $k_z=0$ and $k_z=0.5$. The bulk-paths of $\Gamma-L$, $\Gamma-K$ and $\Gamma-X$ all overlap with the surface k-path $\overline{\Gamma}-\overline{M}$, possibly with contributions from additional paths, such as $X-U$. The plane cuts at different $k_z$  values are derived from the tessellated bulk BZ structure, shown in Figure~\ref{Fig:z_unfolding_illustration}h. When z-unfolding is performed, the value of $k_z$ may be treated as a free parameter. The dependence on $k_z$ manifests as a smooth change in the spectral function over the possible range of $k_z$ which varies the mixture of different constituent bulk-paths that overlap the SBZ-path, as shown in Figure~\ref{Fig:z_unfolding_illustration}i for $\overline{\Gamma}-\overline{M}$. The BZ for z-unfolding is a surface BZ with a finite thickness, shown in red in Figure~\ref{Fig:z_unfolding_illustration}j. The simulation cell for the DFT calculations is set up to be the corresponding real-space unit cell. The z-unfolded k-paths are parallel to the (111) surface at a constant value of $k_z$. 

In ARPES experiments, the relation of the experimental spectra to $k_z$ may be less straightforward. First, the dependence of the inelastic mean free path of the electrons on their kinetic energy is given by the universal curve \cite{2020_Iwasawa_ReviewHighRedARPES_IOP,1979_Seah_UniversalCurveOriginal_SurfInterAna}. Using photon energies that correspond to a small mean free path is advantageous for probing surface states. However, it can produce prominent $k_z$ broadening due to the Heisenberg uncertainty principle \cite{2022_Zhang_ARPESReview_NatureRevMethPrimers,1998_Kumigashira_HighResARPESLaSb_PRB,1974_Feibelman_PhotoemissionSpectQunatumTheoryAndExperiment_PRB, Strocov2003, 2012_Strocov_PRL} that implies integration of the ARPES signal over $k_z$ through the broadening interval. Second, deviations of the photoemission final states from the free electron approximation can cause contributions from different values of $k_z$ to appear in the ARPES spectra. The photoelectrons are often treated as free electrons, based on the assumption that the photoelectron kinetic energy is much larger than the modulations of the crystal potential. In this case, $k_z$ for a given photoelectron kinetic energy, $E_k$, and  the in-plane momentum, $\mathbf{K_{/\!/}}$, is one single value, which is determined by: 
\begin{equation}
{k_{z}=\frac{\sqrt{2m_{0}}}{\hbar}\sqrt{\,E_{k}-\frac{\hbar^{2}}{2m_{0}}K_{/\!/}^{2}-V_{0}}}
\end{equation}
where $m_0$ is the free-electron mass and $V_0$ the inner potential in the crystal. However, a considerable body of evidence has  accumulated that the final states even in metals \cite{1997_Strocov_ExcitedStateCUVLEEDPhotoemission_PRB, 2023_Strocov_HighEnergyPhotoemissionElectronLike_arXiv} and to a greater extent in complex materials such as transition metal dichalcogenides \cite{2006_Strocov_3DBandstructureTiTeFinalState_PRB,2007_Krasovskii_BandMappingPhotoemissionTheory_PRB} can significantly deviate from the free electron approximation. Such deviations can appear, first, as non-parabolic dispersions of the final states and, second, as their multiband composition. The latter means that for given $E_k$ and $\mathbf{K_{/\!/}}$ the final-state wavefunction $\Phi_f$ incorporates a few Bloch waves $\phi_{k_{z}}$ with different $k_z$ values, $\Phi_{f}=\sum_{k_{z}}A_{k_{z}}\phi_{k_{z}}$, which give comparable contributions to the total photocurrent determined by the $A_{k_{z}}$ amplitudes \cite{1997_Strocov_ExcitedStateCUVLEEDPhotoemission_PRB}. A detailed theoretical description of the multiband final states, treated as the time-reversed low-energy electron diffraction (LEED) states \cite{1974_Feibelman_PhotoemissionSpectQunatumTheoryAndExperiment_PRB} within the wavefunction matching approach, as well as further examples for various materials can be found in Refs. \cite{2006_Strocov_3DBandstructureTiTeFinalState_PRB,2007_Krasovskii_BandMappingPhotoemissionTheory_PRB} and the references therein. An insightful analysis of the multiband final states extending into the soft-X-ray photon energies can be found in Ref. \cite{2023_Strocov_HighEnergyPhotoemissionElectronLike_arXiv}. A rigorous analysis of final state effects in ARPES is beyond the scope of this work. Here, we will only mention that all these effects trace back to hybridization of free-electron plane waves through the higher Fourier components of the crystal potential. In cases where significant $k_z$ broadening and/or final states effects are present, z-unfolding, rather than bulk unfolding, should be used in order to resolve the contributions of different $k_z$ values to the measured spectrum. This is demonstrated for CdTe in Section \ref{CdTe}, where the final states appear to incorporate two Bloch waves with $k_z=0$ and $k_z=0.5$.

\subsection{Computational Details}\label{computational}

DFT calculations were conducted using the Vienna Ab Initio Simulation Package (VASP) \cite{1993_Kresse_AbInitioOrigional_PRB} with the projector augmented wave method (PAW) \cite{1994_Blochl_ProjectorAugmentedWaveMethod_PRB,1999_Kresse_UltrasoftPseudopotentialsAumented_PRB}. The generalized gradient approximation (GGA) of Perdew, Burke, and Ernzerhof (PBE) \cite{1996_Perdew_GeneralizedGradientApproximation_PRL} was employed to describe the exchange-correlation interactions among electrons with a Hubbard $U$ correction \cite{1998_Dudarev_LSDAPlusUstudy_PRB}. The $U$ values were machine learned using Bayesian optimization (BO) \cite{2020_YuMarom_BayesianOptimizationHubbardU_npjCompMaterials}. Briefly, the BO objective function is formulated to reproduce as closely as possible the band structure obtained from the Heyd-Scuseria-Ernzerhof (HSE) \cite{2003_Jochen_HSEFunctional_JourChemPhys} hybrid functional. The reference HSE calculations were conducted for bulk CdTe with a lattice parameter of 6.482 \AA {} and $\alpha$-Sn with a lattice parameter of 6.489 \AA {} and compared to the results with the lattice constant of InSb, 6.479 \AA, which was used for interface models. It was verified that using the lattice constant of InSb does not have an appreciable effect on the electronic properties of CdTe and $\alpha$-Sn, as shown in the SI.

The hyperparameters of our BO implementation are the coefficients $\alpha_1$ and $\alpha_2$, which assign different weights to the band gap vs. the band structure in the objective function, the number of valence and conduction bands used for the calculation of the objective function, $N_b$, and the parameter $\kappa$ that controls the balance between exploration and exploitation in the upper confidence bound acquisition function.  For InSb the values of $U^{In,p}_{eff}=-0.2$ and $U^{Sb,p}_{eff}=-6.1$ were used, following Ref.~\cite{2022_Yang_TransforamtionMatrixBandUnfoldingInAsInSbSurfaces_AdvQuanTech,2020_YuMarom_BayesianOptimizationHubbardU_npjCompMaterials}. It has been shown that PBE+U(BO) produces a band structure in good agreement with ARPES for InSb ~\cite{2022_Yang_TransforamtionMatrixBandUnfoldingInAsInSbSurfaces_AdvQuanTech}.

Because $\alpha$-Sn is a semi-metal, only the band shape was considered in the optimization, \textit{i.e.} $\alpha_1$ was set to 0 and $\alpha_2=1$ \cite{2022_Dardzinski_BestPracticesDFTInorganicInterfacesBO_JPHYS-CONDENSMAT}.The other BO hyperparameters used for Sn were $\kappa=7.5$ and $N_b=(5,5)$. This resulted in a value of $U^{Sn,p}_{eff}=-3.04$ eV, slightly different than in Refs. \cite{2014_Kufner_TopoaSnSurfaceStatesVsThicknessStrain_PRB,2020_Shi_FirstprincipleTopologicalSurfaceaSn111_Elsevier.pdf,2013_Barfuss_TopoInsulatoraSnInSb001_PRL}, which used empirical methods to choose a $U$ value that yields a correct band ordering. As shown in Ref. \cite{2022_Dardzinski_BestPracticesDFTInorganicInterfacesBO_JPHYS-CONDENSMAT}, PBE+U(BO) reproduces the correct band ordering of $\alpha$-Sn with the band inversion at the $\Gamma$ point, in agreement with other studies using DFT+U \cite{2022_InSbaSn001ARPESCalibrationTicknessDependent_PRB,2017_Rogalev_DoubleBandInversionaSnOrbital_PRB}.

For CdTe, we applied a $U$ correction to both the Cd-$d$ orbitals and Te-$p$ orbitals, unlike earlier studies \cite{2021_Yang_FirtsPrincipleAssessInAs-GaSbInterfaceCdTe_PRM.pdf,2015_Wu_CdTeLDAUParameter.pdf}. The hyperparameters used for CdTe were $\kappa=7.5$, $N_b=(5,5)$, $\alpha_1=0.5$ and $\alpha_2=0.5$. The latter two parameters were chosen to assign equal weights to the band gap and the band shape. This led to $U$ values of $U^{Cd,d}_{eff}=7.381$ and $U^{Te,p}_{eff}=-7.912$. The Cd-$d$ $U$ value obtained here is similar to the 7 eV used in Ref.~\cite{2015_Wu_CdTeLDAUParameter.pdf} and somewhat lower than $U^{Cd,d}_{eff}=8.3$ eV in Ref.~\cite{2021_Yang_FirtsPrincipleAssessInAs-GaSbInterfaceCdTe_PRM.pdf}. The gap of 1.21 eV, obtained here by applying the Hubbard $U$ correction to both the Te-$p$ states and the Cd-$d$ states is closer to experimental values of around 1.5 eV \cite{2000_Fonthal_TempDependenceBandGapCdTe,2014_Hinuma_BandAlignmentDFTInSbCdTeInterfaces_PRB} and the HSE value of 1.31 eV than previous calculations  \cite{2021_Yang_FirtsPrincipleAssessInAs-GaSbInterfaceCdTe_PRM.pdf}.

Spin-orbit coupling (SOC) was used in all calculations and dipole corrections were applied to slab models \cite{1992_Neugebauer_VASPDipoleCorrectionImplement_PRB}. The tags used for convergence of calculations were BMIX = 3, AMIN = 0.01, ALGO = Fast, and EDIFF = 1$\cdot 10^{-5}$. The kinetic energy cutoff was set to 400 eV for all bulk calculations and 350 eV for surface and interface slab models. A  $9\times 9\times 9$ k-point mesh was used for  bulk calculations  and a k-point mesh of $7\times 7\times 1$ was used for surface and interface calculations. All interface density of states (DOS) calculations used a k-point mesh of $13\times 13\times 1$. 

All band structure and density of states plots were generated using the open-source Python package, VaspVis \cite{2022_Dardzinski_BestPracticesDFTInorganicInterfacesBO_JPHYS-CONDENSMAT}, which is freely available from The Python Package Index (PyPI) via the command: \textit{pip install vaspvis}, or on GitHub at: \url{https://github.com/DerekDardzinski/vaspvis}

\subsection{Slab Construction}
All slab models were constructed using the experimental InSb lattice constant value of 6.479 \AA {} \cite{2001_Vurgaftman_InSbIIIVSemiconductorsProperties_JAPPLPHYS}, assuming that the epitaxial films of CdTe and $\alpha$-Sn would conform to the substrate. The length of two monolayers of a (110) slab was 4.5815 \AA in the z-direction. A vacuum region of around 40 \AA {} was added to each slab model in the z-direction to avoid spurious interactions between periodic replicas. The surfaces of all slab models were passivated by pseudo-hydrogen atoms such that there were no surface states from dangling bonds \cite{2005_Huang_SurfacePassivationMethodNanostructures}. Despite $\alpha$-Sn being a semi-metal passivation is required to remove spurious surface states, as shown in the supplemental information (SI). The pseudo-hydrogen fractional charges utilized to passivate each atom were 1.25 for In and 0.75 for Sb in InSb, 1.5 for Cd and 0.5 for Te in CdTe, and 1 for Sn. Structural relaxation of the pseudo-hydrogen atoms was performed until the maximal force  was below 0.001 eV/\AA. The InSb/CdTe interface structure has In-Te and Sb-Cd bonds with each In interface atom connected to 3 Sb and 1 Te. The configuration with In-Cd and Sb-Te bonds was also considered but this was found to be less stable by 1.33 eV. Ideal interfaces were considered with no intermixing and no relaxation of the interface atoms was performed. 

When constructing such slab models, it is necessary to converge the number of layers to avoid quantum size effects and approach the bulk properties  \cite{2020_Yang_TopoInterfacesWavefunctionDensity_PRM}. For InSb it has previously been shown that 42 monolayers are sufficiently converged \cite{2022_Yang_TransforamtionMatrixBandUnfoldingInAsInSbSurfaces_AdvQuanTech}. Plots of the band gap vs. the number of atomic layers for CdTe(110) and $\alpha$-Sn (110) slabs are provided in the SI. CdTe was deemed converged with 42 monolayers with a gap value of 1.23 eV, which is only slightly larger than the bulk PBE+U(BO) value. The z-unfolded band structures of CdTe(111) were calculated for a 40 monolayer slab.  A 26 monolayer slab model was used to simulate the 2 $\times$ 2 reconstruction, due to the higher computational cost of the 2 $\times$ 2 supercell. Structural relaxation was performed for the top two monolayers of the 2 $\times$ 2 reconstruction. For the slab of unstrained (110) $\alpha$-Sn, 70 monolayers were needed to close the gap at the zero-gap point of the semi-metal, which  corresponds to around 16 nm. The tri-layer slab models comprised 42 layers of InSb, 70 layers of $\alpha$-Sn and between 0 and 16 layers of CdTe in two-layer increments, amounting to a total slab thickness of around 300 nm (not including vacuum). The (110) bi-layer slab models comprised 42 layers of CdTe and InSb, and 70 layers of $\alpha$-Sn as these were deemed converged.

\subsection{ARPES Experimental details}
The $\alpha$-Sn samples were grown by molecular beam epitaxy on an In-terminated c(8 $\times$ 2) InSb(001) surface prepared by atomic hydrogen cleaning. 51 monolayers (16.5 nm) of $\alpha$-Sn were deposited as calibrated via Rutherford backscattering spectrometry. Growth was performed at a substrate temperature of -20 $^\circ$C and a base pressure better than 1$\cdot 10^{-10}$ Torr. The ARPES measurements were taken at Beamline 10.0.1.2 at the Advanced Light Source in Berkeley. The base pressure was better than 5$\cdot 10^{-11}$ Torr while the sample temperature was held at 68 K. The sample was illuminated with 63 eV \textit{p}-polarized light and spectra were collected using a Scienta R4000 detector with energy resolution better than 40 meV and angular resolution better than 0.1$^\circ$. The sample was transferred via vacuum suitcase with a base pressure better than $\cdot 10^{-11}$ Torr  between the growth chamber and beamline. A photon energy of 63 eV corresponds to a $\mathbf{k}_z$  approximately 0.15 \AA$^{-1}$ above the $\mathbf{\Gamma}_{002}$ point.

\section{Results and discussion\label{Sec:Results}}

\subsection{$\alpha$-Sn}\label{Sn}

\begin{figure}[h]
\centering
\includegraphics[width=0.47\textwidth]{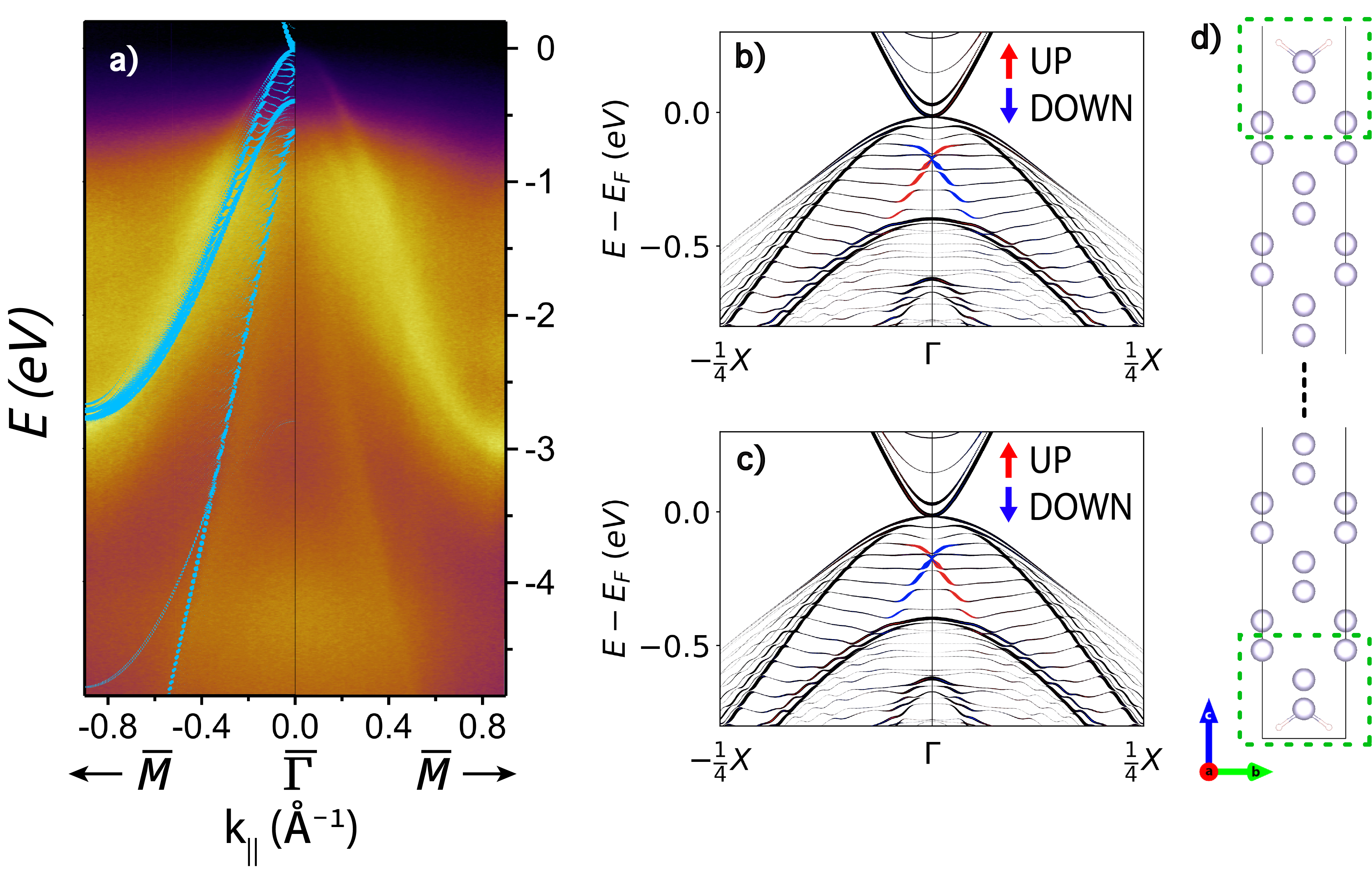} 
\caption{Electronic structure of $\alpha$-Sn:
(a) Bulk-unfolded band structure of an $\alpha$-Sn (001) slab with 51 atomic layers (light blue) compared with ARPES data for a sample of the same thickness. The point $\overline{M}$ is at $0.9298$ \AA$^{-1}$. The ARPES data is cutoff at $0.9$ \AA$^{-1}$ due to experimental artifacts at the edges. Spin-polarized band structures projected onto (b) the top surface atoms and (c) the bottom surface atoms, indicated by the green boxes on the slab structure illustrated in (d).
\label{Fig:aSn_ARPES_DFT_MGM}}
\end{figure}

Figure \ref{Fig:aSn_ARPES_DFT_MGM}a shows the bulk unfolded PBE+U(BO) band structure for a 51 monolayer thick $\alpha$-Sn (001) slab, compared to  ARPES data for a sample of the same thickness taken at a photon energy of 63 $eV$. The point $\overline{M}$ is at $0.9298$ \AA$^{-1}$. The ARPES data is cutoff at $0.9$ \AA$^{-1}$ due to experimental artifacts at the edges. The PBE+U (BO) band structure is in excellent agreement with ARPES. The top of the valence band in the ARPES and the simulated band structure lines up and the bulk bands are reproduced well. The bandwidth of the heavy hole band, $\Gamma_8$, is  slightly underestimated, consistent with Ref. ~\cite{2022_Yang_TransforamtionMatrixBandUnfoldingInAsInSbSurfaces_AdvQuanTech}. This is corrected by the HSE functional, as shown in the SI for a bulk unit cell of $\alpha$-Sn with a (001) orientation. However, it is not feasible to use HSE for the large interface models studied here, owing to its high computational cost.

The previously reported topological properties of $\alpha$-Sn slabs are also observed here \cite{2017_Rogalev_DoubleBandInversionaSnOrbital_PRB,2022_InSbaSn001ARPESCalibrationTicknessDependent_PRB,2020_Shi_FirstprincipleTopologicalSurfaceaSn111_Elsevier.pdf,2017_Xu_TopoDiracSemimetalaSnInSb111_PRL,2013_Barfuss_TopoInsulatoraSnInSb001_PRL,2018_Scholz_TopoSurfaceStateaSnPhotoemission_PRB,2017_Huang_StrainedSnDiracSemimetalMagnetoresitance_PRB,2019_Rogalev_TailoringTopoSurfaceStateaSnFilms_PRB}. The spin-polarized topological surface state (TSS) is shown in panels (b) and (c) of Fig.~\ref{Fig:aSn_ARPES_DFT_MGM}  for a (001) 51 monolayer slab along the $X-\Gamma-X$ k-path. As expected, the TSS is characterized by a linear dispersion with the top and bottom surfaces having opposite spin polarization. The associated Rashba-like surface states are also observed along the $K-\Gamma-K$ k-path, as shown in the SI. This linear surface state is also observed in the (110) slabs used to construct the bilayer and tri-layer models. Notably there is an energy gap between the top and bottom TSSs, which closes at 70 layers, the same thickness at which the band gap closes. This gap is possibly induced by the hybridization of the top and bottom surface states in under-converged slabs. We note that the effect of strain on the electronic structure of $\alpha$-Sn is not studied here.

\subsection{CdTe}\label{CdTe}

\begin{figure*}[t]
\centering
\includegraphics[width=1\textwidth]{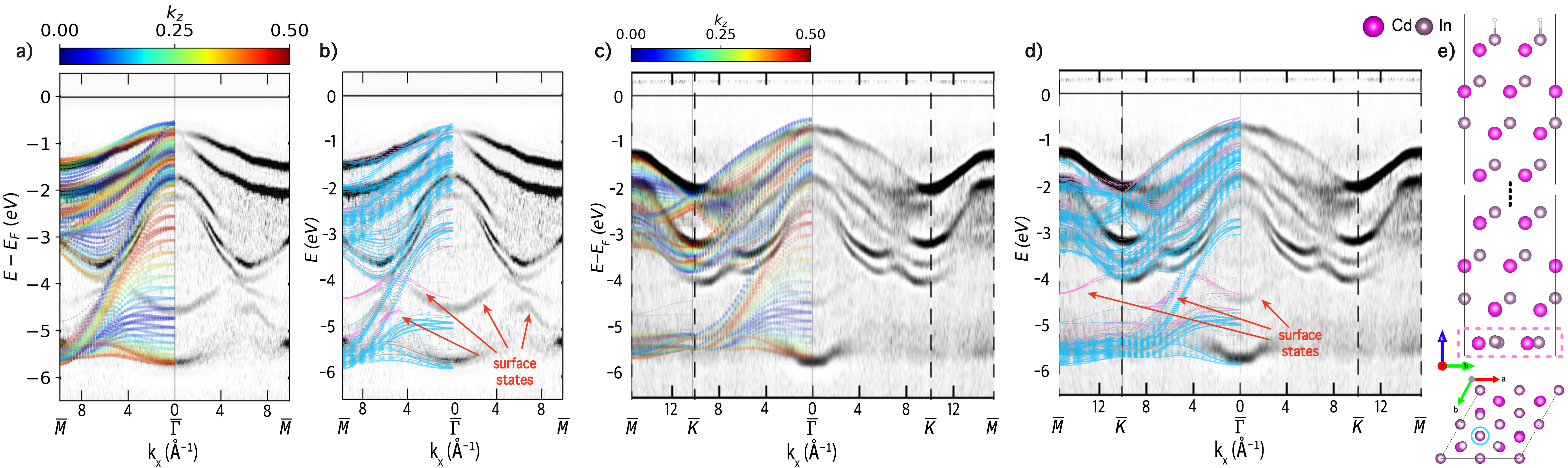}
\caption{Electronic structure of CdTe: Z-unfolded band structures of CdTe(111) compared with second-derivative map of ARPES data (black and white), adapted with permission from “Spectroscopic studies of CdTe(111) bulk and surface electronic structure” by J. Ren \textit{et al.}, Phys. Rev. B, 91, 235303 (2015); Copyright (2015) by the American Physical Society \cite{2015_Ren_CdTe111ARPESMGM_PRB}.  Z-unfolded band structures compared to ARPES data along (a), (b) $\overline{\Gamma}-\overline{M}$ and (c), (d) $\overline{\Gamma}-\overline{K}-\overline{M}$.  (a), (c) Dependence of the band structure on $k_z$. (b), (d) Mixture of $k_z=0.0$ and $k_z=0.5$ (cyan) for a model with a 2 $\times$ 2 surface reconstruction with the contributions of the surface atoms shown in pink. DFT has shift of -0.25 $eV$ and stretch factor of 1.22 for comparison. (e) Illustration of the 2 $\times$ 2 surface reconstruction with the Cd atom removed indicated by a blue circle. The atoms used for the surface projection are indicated by a pink dashed box}
\label{Fig:CdTe_zunfolding_ARPES_BothPaths}
\end{figure*}

Fig.~\ref{Fig:CdTe_zunfolding_ARPES_BothPaths} shows a comparison of band structures obtained using PBE+U(BO) to the ARPES experiments of Ren \textit{et al.} \cite{2015_Ren_CdTe111ARPESMGM_PRB} for CdTe(111). Ren \textit{et al.} collected ARPES data at photon energies of 19, 25 and 30 $eV$. Here, we compare our results with the second-derivative maps of the ARPES data taken at 25 $eV$ along the k-paths $\overline{\Gamma}-\overline{M}$ (panels (a) and (b)) and $\overline{\Gamma}-\overline{K}-\overline{M}$ (panels (c) and (d)). The original data has been converted to gray scale and reflected around $k_x=0$. To facilitate the qualitative comparison of the DFT band structure features with the ARPES experiment, we apply a Fermi energy shift of 0.25 $eV$ to line up the VBM and a stretch factor of 1.22 to compensate for the bandwidth underestimation of PBE+U(BO), particularly for bands deep below the Fermi energy \cite{2014_Kronik_ShiftAndStretchSectionPg163_BookSpringer}. Bandwidth underestimation by PBE+U(BO) compared with HSE and ARPES has also been reported for InAs and InSb in  \cite{2022_Yang_TransforamtionMatrixBandUnfoldingInAsInSbSurfaces_AdvQuanTech,2006_Krukau_PBEHSECompare_JourChemPhys}. The original computed band structure without the shift and stretch is provided in the SI.  

Owing to the low mean free path at this photon energy, the spectrum appears integrated over a certain $k_z$ interval and surface contributions are readily visible in the ARPES \cite{2020_Iwasawa_ReviewHighRedARPES_IOP,1979_Seah_UniversalCurveOriginal_SurfInterAna}. To account for the different $k_z$ contributions, the z-unfolding method was employed, as described in Section \ref{Z-Unfolding}. Panels (a) and (c) show the z-unfolded band structures as a function of $k_z$ for slab models without a surface reconstruction (figures with single values of $k_z$ are provided in the SI). This is used determine which $k_z$ values are likely present in the experiment. A mixture of $k_z=0$ and $k=0.5$ provides the best agreement with the ARPES data. This combination of $k_z$ values is used for the DFT data shown in cyan in panels (b) and (d). This is consistent with the $k_z$ broadening with contributions centered around $k_z=0$ and $k=0.5$ often present in ARPES data taken at low mean field path energies in gapped materials \cite{2022_Zhang_ARPESReview_NatureRevMethPrimers,1998_Kumigashira_HighResARPESLaSb_PRB,1974_Feibelman_PhotoemissionSpectQunatumTheoryAndExperiment_PRB_PhotoemissionSpectQunatumTheoryAndExperiment_PRB}.

To account for the presence  of surface states, we modeled the CdTe(111)A-(2 $\times$ 2) surface reconstruction \cite{2010_Egan_2X2CdTeExperiment_JVSTA}, illustrated in panel (e). The atom-projected band structures of the bottom layer (indicated by pink dashed box) are plotted in pink in panels (b) and (d). The additional bands arising from the surface reconstruction are in close agreement with the bands in the ARPES labeled as surface states by Ren \textit{et al.}, indicated by red arrows. These surface states are unaffected by the choice of $k_z$. By accounting for the contributions of different $k_z$ values and for the presence of surface states excellent agreement with experiment is achieved, as the DFT band structures reproduce all the features of the ARPES.

%%%%%%%%%%%%%%%%%%%%%%%%%%%%%%%%%%%%%%%%%%%%%%%%%%%%%%%%%%%%%%%%%%%%%%%%%%%%%%%%%%%%%%%%%%%%%%%%%%%%%%%%%%%%%%%%%%%%%%%%%%%%%
 \subsection{Bilayer Interfaces}\label{bilayer}

\begin{figure*}[t]
\centering
\includegraphics[width=1\textwidth]{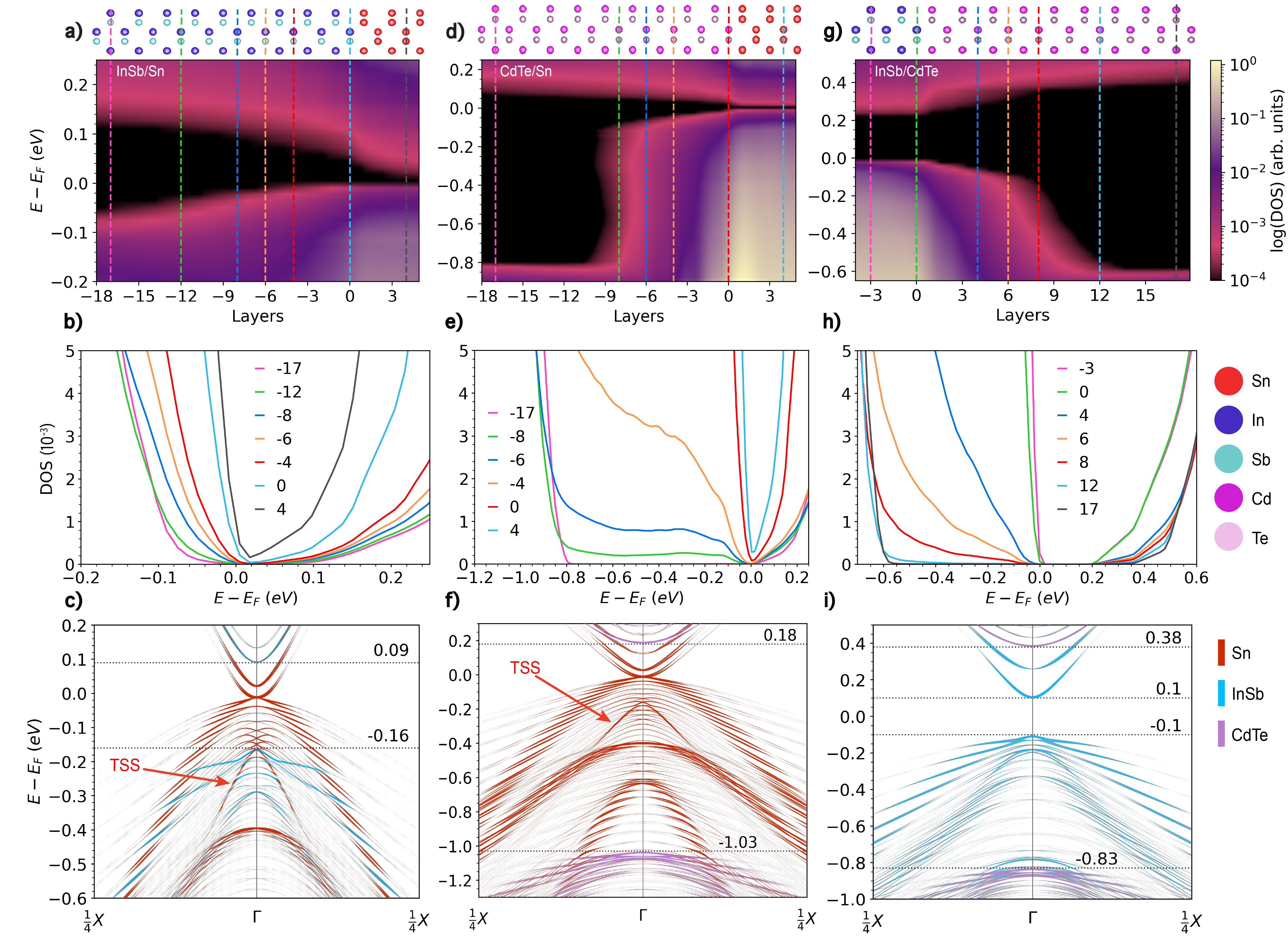}
\caption{Electronic structure of bilayer interfaces: Density of states in the (a) InSb/$\alpha$-Sn, (d) CdTe/$\alpha$-Sn and (g) InSb/CdTe interfaces as a function of position. The atomic layers are numbered based on distance from the interface, which is located at zero. The structure of each interface is illustrated on top. (b Local density of states for selected layers in the (b) InSb/$\alpha$-Sn, (e) CdTe/$\alpha$-Sn and (h) InSb/CdTe interfaces, indicated by dashed lines in the same colors in panels (a), (d), and (g), respectively. Element projected band structures of the (c) InSb/$\alpha$-Sn, (f) CdTe/$\alpha$-Sn and (i) InSb/CdTe interfaces, with bands originating from $\alpha$-Sn colored in red,  bands originating from InSb colored in light blue, and  bands originating from CdTe colored in purple. 
\label{Fig:BiLayersDOS}}
\end{figure*}

We begin by probing the local electronic structure at the the InSb/$\alpha$-Sn bi-layer interface. Fig.~\ref{Fig:BiLayersDOS}a shows the DOS as a function of position across the interface, indicated by the atomic layer number. Fig.~\ref{Fig:BiLayersDOS}b shows the local DOS at select positions. The Fermi level is positioned at the semi-metal point of the $\alpha$-Sn and in the gap of the InSb. We note that the $\alpha$-Sn appears as if it has a small gap due to an artifact of the $10^{-4}$ cutoff applied in the log plot in panels (a) and (d). The local DOS plots shown in panels (b) and (e) and the band structure plots shown in panels (c) and (f) clearly show the semi-metal point. No significant band bending is found for InSb, as expected from branching point theory \cite{tersoff1986calculation, 1996_Monch_JApplPhys}. Based on the element-projected band structure, shown in panel (c), the InSb conduction band minimum (CBM) lies 0.09 eV above the $\alpha$-Sn semi-metal point and the InSb valence band maximum (VBM) lies 0.16 eV below it. A linear TSS is present in the $\alpha$-Sn. Based on an atom projected band structure, shown in the SI, the origin of this state is the top surface of $\alpha$-Sn, adjacent to the vacuum region.  A TSS is no longer present in the $\alpha$-Sn layers at the interface with InSb, possibly owing to hybridization between the $\alpha$-Sn and InSb \cite{2019_Rogalev_TailoringTopoSurfaceStateaSnFilms_PRB}. Metal-induced gap states (MIGS) are an inherent property of a metal/semiconductor interface, produced by the penetration of exponentially decaying metallic Bloch states into the gap of the semiconductor  \cite{1987_MIGSVGSProgenitorWork_PRL, 1999_Monch_SchottkyContactsExplainedMIGSInterfaces_AIP, 1990_Monch_PhysicsofMetalSemiconductorInterfaces_RepProgPhysBook,2013_Robertson_BandOffsetsSchottkyBarrierMIGS_JVSTA}. The presence of MIGS manifests in Figure Fig. ~\ref{Fig:BiLayersDOS}a  as a gradually decaying non-zero DOS in the band gap of the InSb in the vicinity of the interface. Figure ~\ref{Fig:BiLayersDOS}b shows that the MIGS are prominent in the first few atomic layers and become negligible beyond 8 layers from the interface.

Fig.~\ref{Fig:BiLayersDOS}d shows the DOS as a function of position across the CdTe/$\alpha$-Sn interface, indicated by the atomic layer number. Fig.~\ref{Fig:BiLayersDOS}e shows the local DOS at select positions.  The Fermi level is positioned at the semi-metal point of the $\alpha$-Sn and in the gap of the CdTe. Based on the projected band structure, shown in panel (f), the CdTe CBM is positioned 0.18 eV above the Fermi level and the CdTe VBM is located 1.03 eV below the Fermi level. This agrees with previous reports that interfacing with Sn brings the conduction band of the CdTe closer to the Fermi energy, with downward band-bending of 0.25 $eV$ \cite{1987_Tang_ARPESaSnCdTe100Offset_PRB} and 0.1 $eV$ \cite{1985_Takatani_ThinSnCdTe111_PRB}. We find a valence band offset of around 1 eV, similar to the (110) and (111) interface reported in  \cite{1988_Hochst_aSnCdTe110Offset_JVSTB,1990_Lambrecht_InterfacePOlaritySemiconductroHetrojunctionBandOffset_PRB,1985_Takatani_ThinSnCdTe111_PRB,1992_ContinenzaaSnCdTeOrientationsBandLineupStudy_PRB,2018_Coster_HamiltonianStudyaSnonCdTeThinFilms_PRB,1987_Cardona_HetrostuctureBandOffsets_PRB}. Close to the interface there is a significant density of MIGS, which decay within about 10 layers (3-4 nm) into the CdTe. This suggests that this number of CdTe layers may be required for an effective tunnel barrier.

Fig.~\ref{Fig:BiLayersDOS}g shows the DOS as a function of position across the InSb/CdTe interface, indicated by the atomic layer number. Fig.~\ref{Fig:BiLayersDOS}h shows the local DOS at select positions. The band alignment is type-I with the CdTe band gap straddling the InSb band-edges. The Fermi level is close to the InSb VBM and around the middle of the gap of the CdTe. No band bending is found in either material. Based on the  projected band structure, shown in panel (i), the CdTe CBM lies 0.28 eV above the InSb CBM and the CdTe VBM lies 0.75 eV below the InSb VBM. These values are similar to the band offsets reported in references ~\cite{2022_Badawy_EpitaxyCdTeOnInSbNanowires_ADVSCI, 2014_Hinuma_BandAlignmentDFTInSbCdTeInterfaces_PRB,Wang_2018_BandAlignmentCdTeInSb001_JourVacSTA}. Because the band gap of InSb is significantly smaller than that of CdTe, states from the InSb penetrate into the gap of the CdTe, similar to MIGS. These states decay gradually and vanish at a distance greater than 12 layers from the interface.

\subsection{Tri-layer Interfaces}\label{tri-layer}

\begin{figure*}[t]
\centering
\includegraphics[width=1\textwidth]{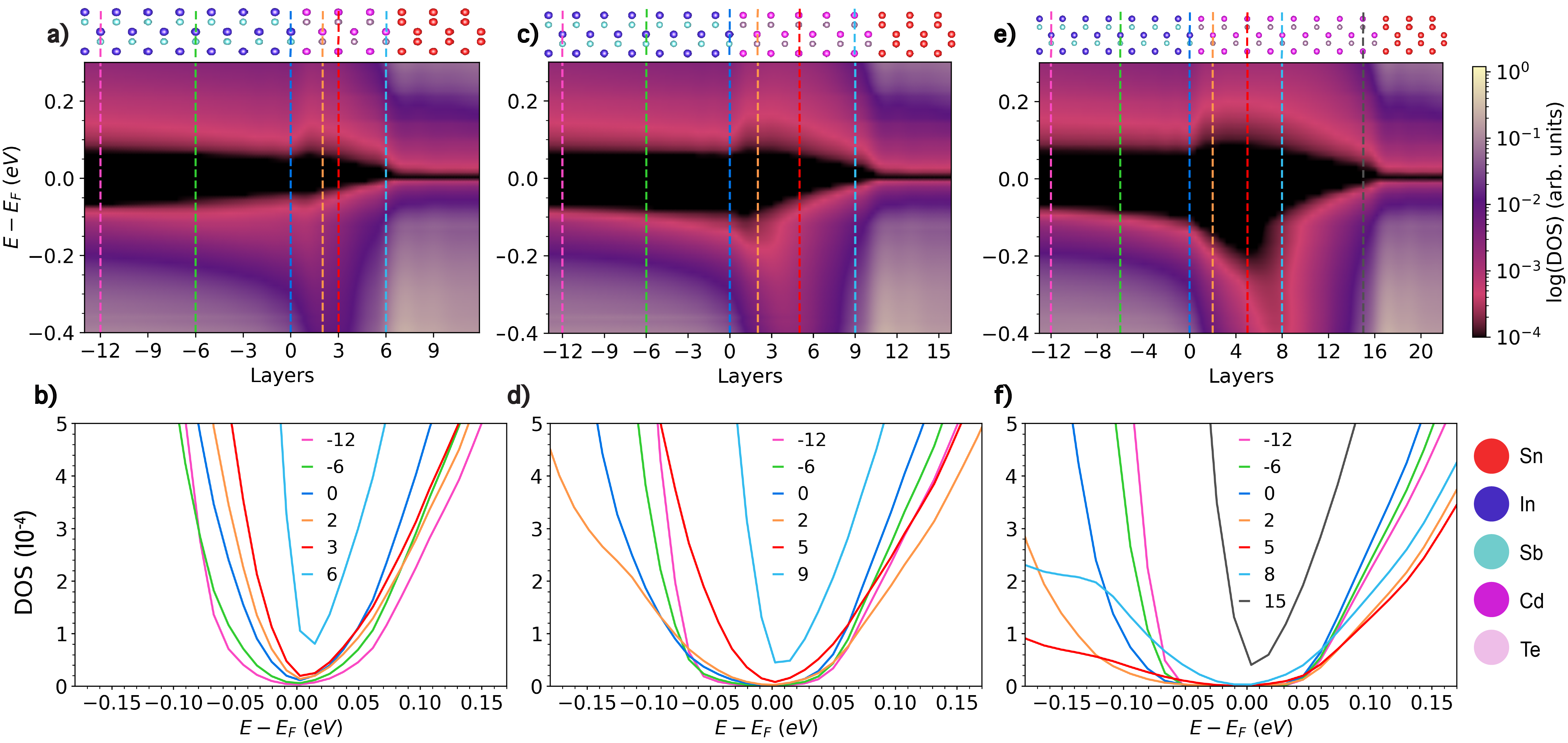}
\caption{
Electronic structure of InSb/CdTe/$\alpha$-Sn tri-layer interfaces: Density of states as a function of distance from the interface for (a) 6, (c) 10 and (e) 16 CdTe barrier layers. 
The atomic layers are numbered based on distance from the interface, which is located at zero. Interface structures are illustrated on top. 
(b), (d), (f) Local density of states for selected layers, indicated by dashed lines in the same colors in panels (a), (c), and (e), respectively. 
\label{Fig:Tri-layerDOS}}
\end{figure*}

Fig.~\ref{Fig:Tri-layerDOS} shows the DOS as a function of position across InSb/CdTe/$\alpha$-Sn tri-layer interfaces with varying thickness of the CdTe tunnel barrier. The position is indicated by the atomic layer number, with the layer of InSb closest to the CdTe considered as zero. Panels (a) and (b) show that with 6 atomic layers of CdTe, the MIGS from the $\alpha$-Sn penetrate through the tunnel barrier into the first 12 layers of the InSb. For a thin layer of CdTe, the band gap is expected to be significantly larger than the bulk value because of the quantum size effect (see the gap convergence plot in the SI). However, owing to the presence of MIGS, the gap of the CdTe remains considerably smaller than its bulk value. With 10 layers of CdTe, shown in panels (c) and (d), there is still a significant presence of MIGS throughout the CdTe, which decay by 6 layers into the InSb. Panels (e) and (f) show that with 16 layers of CdTe the InSb is completely insulated from MIGS coming from the $\alpha$-Sn. The gap of the CdTe reaches a maximum of around 0.3 $eV$ at a distance of 5 layers from the InSb. This is because MIGS from the $\alpha$-Sn penetrate into the CdTe from one side, whereas states from the InSb penetrate from the other side, such that the band gap of the CdTe never reaches its expected value.

\begin{figure}[h]
\centering
\includegraphics[width=0.45\textwidth]{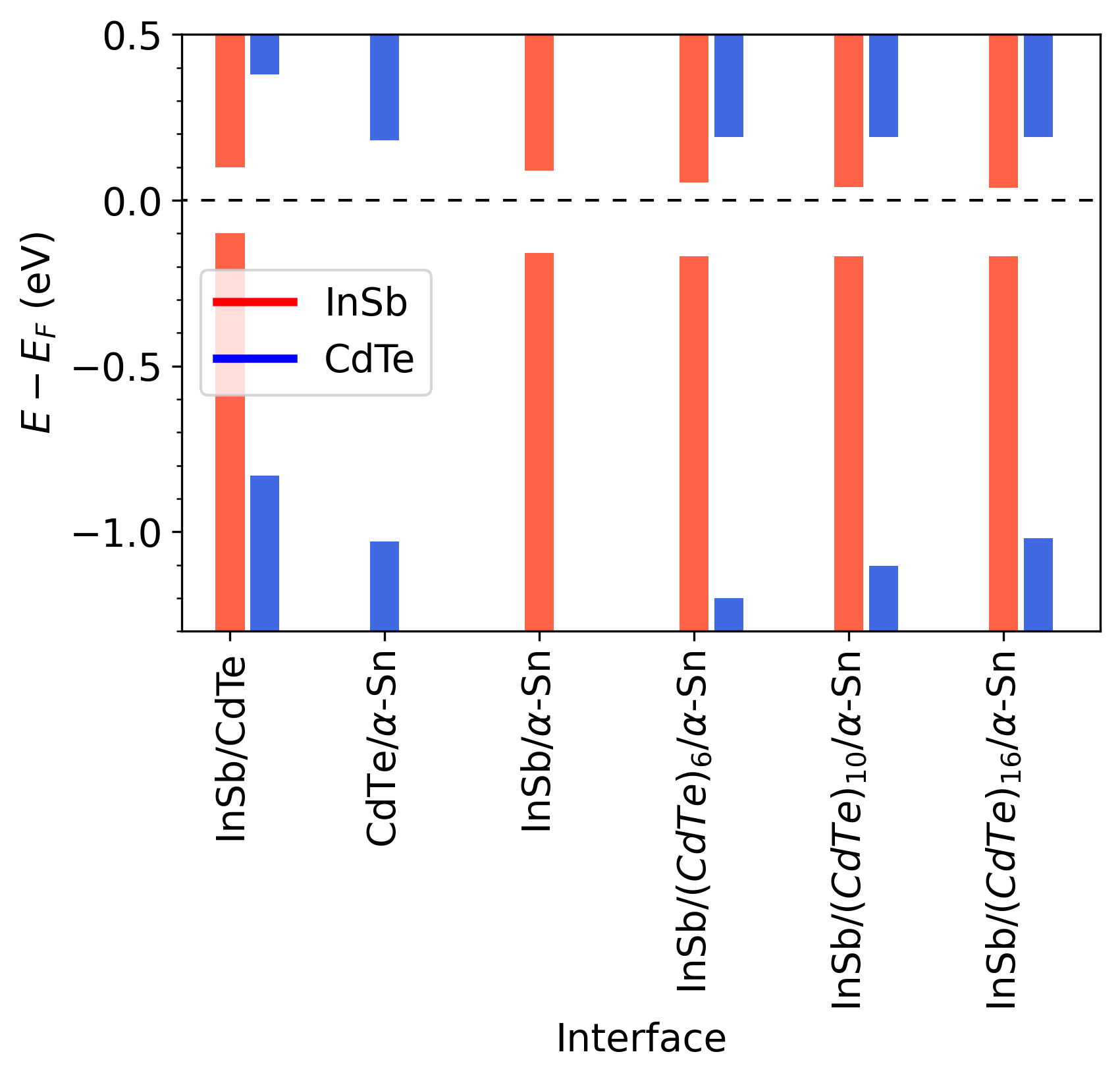} 
\caption{Valence and conduction band edge positions for InSb and CdTe in the bilayer and tri-layer interfaces. The Fermi level is at the semi-metal point of the $\alpha$-Sn.
\label{Fig:band_edge_pos}}
\end{figure}

Figure \ref{Fig:band_edge_pos} summarizes the band alignment at the bilayer and tri-layer interfaces studied here. For the tri-layer interfaces, the band alignment between the InSb and the $\alpha$-Sn is not significantly affected by the presence of CdTe, as shown in the element-projected band structures in the SI. The $\alpha$-Sn semi-metal point remains pinned at the Fermi level, as in the bilayer InSb/$\alpha$-Sn (see also Figure \ref{Fig:BiLayersDOS}c). The InSb VBM remains at 0.17 eV below the Fermi level, similar to its position in the bilayer interface, regardless of the CdTe thickness. The InSb CBM position decreases slightly with the thickness of the CdTe from 0.09 eV above the Fermi level without CdTe, to 0.054 eV with 6 layers of CdTe, 0.04 eV with 10 layers, and 0.037 eV with 16 layers. This may be attributed to the quantum size effect, which causes a slight narrowing of the InSb gap because of the increase in the overall size of the system. Based on the element-projected band structures provided in the SI, the band edge positions of the CdTe are dominated by the interface with the $\alpha$-Sn, rather than the interface with the InSb. The CdTe CBM remains at 0.18 eV above the Fermi level, as in the bilayer CdTe/$\alpha$-Sn interface (see also Figure \ref{Fig:BiLayersDOS}f), regardless of the number of layers. As the band gap of the CdTe narrows with increasing thickness, the CdTe VBM shifts from 1.24 eV below the Fermi level with 6 layers to 1.105 eV with 10 layers, and 1.05 eV with 16 layers, approaching the bilayer VBM position of 1.03 eV below the Fermi level with 42 layers. Although the band gap of the CdTe is significantly reduced due to MIGS, a type I band alignment with the InSb is maintained, similar to the bilayer InSb/CdTe interface (Figure \ref{Fig:BiLayersDOS}g,i), as shown in Fig.~\ref{Fig:Tri-layerDOS} panels (a), (c), and (e).

\begin{figure}[h]
\centering
\includegraphics[width=0.45\textwidth]{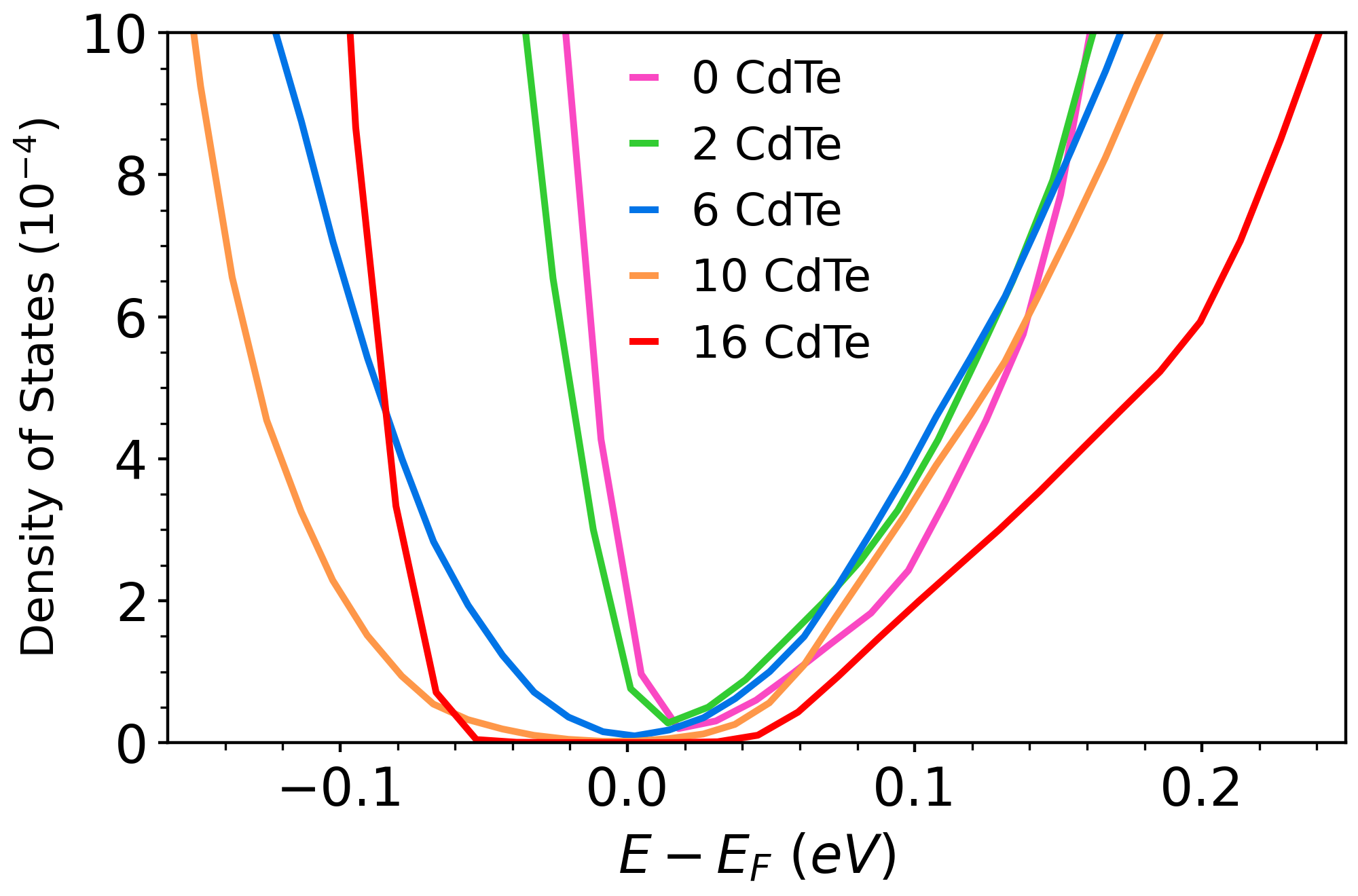} 
\caption{Density of states in the second InSb layer  from the interface (layer -2 in Figure \ref{Fig:Tri-layerDOS}) as a function of the number of CdTe barrier layers.
\label{Fig:multiple_DOS_cuts}}
\end{figure}

Figure \ref{Fig:multiple_DOS_cuts} show the LDOS in the second layer of InSb from the interface as a function of the number of CdTe layers. Without CdTe and with two layers of CdTe, there is no band gap in the InSb close to the interface, owing to the significant density of MIGS. With 6 layers of CdTe the gap of the InSb close to the interface is still considerably narrower than its bulk value. The band gap in the second layer of InSb from the interface approaches its bulk value with 10 layers of CdTe and finally reaches it with 16 layers of CdTe. This suggests that 16 CdTe layers provide an effective barrier to electronically insulate the InSb from the $\alpha$-Sn.
It is reasonable to assume that a barrier of this thickness or higher would all but eliminate transport through the interface into the InSb. Therefore, we estimate that the relevant barrier thickness regime to modulate the coupling at an interface with a superconductor and tune the proximity effect would be in the range of 6-10 layers, where MIGS still exist. We note, however, that the interface with $\beta$-Sn may have somewhat different characteristics in terms of the band alignment and the penetration depth of MIGS.

\section{Conclusion}

In summary, we have used DFT with a Hubbard U correction machine-learned by Bayesian optimization to study CdTe as a prospective tunnel barrier at the InSb/$\alpha$-Sn interface. The results of PBE+U(BO) were validated by comparing the band structures of slab models of $\alpha$-Sn(001) and CdTe(111) with ARPES experiments (the PBE+U(BO) band structure of InSb(110) had been compared to ARPES experiments previously \cite{2022_Yang_TransforamtionMatrixBandUnfoldingInAsInSbSurfaces_AdvQuanTech}). Excellent agreement with experiment is obtained for both materials. In particular, for the low-mean-free-path ARPES of CdTe, the z-unfolding scheme successfully reproduces the contributions of different $k_z$ values and modelling the 2 $\times$ 2 surface reconstruction successfully reproduces the contributions of surface states. 

We then proceeded to use PBE+U(BO) to calculate the electronic structure of bilayer InSb/$\alpha$-Sn, CdTe/$\alpha$-Sn, and InSb/CdTe, as well as tri-layer InSb/CdTe/$\alpha$-Sn interfaces with varying thickness of CdTe. Simulations of these very large interface models were possible thanks to the balance between accuracy and computational cost provided by PBE+U(BO). We find that the most stable configuration of the InSb/CdTe interface is with In-Te and Sb-Cd bonding. MIGS penetrate from the $\alpha$-Sn into the InSn and CdTe. Similarly, states from the band edges of InSb penetrate into the larger gap of the CdTe. No interface states are found in any of the interfaces studied here, in contrast to the EuS/InAs interface, for example, in which a quantum well interface state emerges \cite{2021_Yu_EuSInSbBondingConfig_PRM}. 

For all interfaces comprising $\alpha$-Sn, the semi-metal point is pinned at the Fermi level. For the tri-layer interface, the band alignment between the InSb and the $\alpha$-Sn remains the same as in the bilayer interface regardless of the thickness of the CdTe barrier, with the Fermi level closer to the conduction band edge of the InSb. The band edge positions of the CdTe are dominated by the interface with the $\alpha$-Sn rather than the interface with InSb, with the conduction band edge being closer to the Fermi level. A type-I band alignment is maintained between CdTe and InSb with the gap of the former straddling the latter. The CBM of the CdTe is pinned whereas the VBM shifts upwards towards the Fermi level as the gap narrows with the increase in thickness.   

We find that 16 layers of CdTe (about 3.5 nm) serve as an effective barrier, preventing the penetration of MIGS from the $\alpha$-Sn into the InSb. However, in the context of Majorana experiments, it is possible that a barrier thick enough to completely insulate the semiconductor from the superconductor would also all but eliminate transport. Therefore, we estimate that the relevant regime for tuning the coupling at the interface would be in the thickness range where some MIGS are still present, while thicker CdTe layers could be used to passivate exposed InSb surfaces. We note, however, that the interface with the superconducting $\beta$-Sn, which is not lattice matched to InSb and CdTe, may have different characteristics than the interface with $\alpha$-Sn. In practice, careful experimentation with varying barrier thickness would be needed to determine the optimal configuration for MZM devices.     

We have thus demonstrated that DFT simulations can provide useful insight into the electronic properties of semiconductor/tunnel barrier/metal interfaces. This includes the interface bonding configuration, the band alignment, and the presence of MIGS (and, possibly, of interface states). Such simulations may be conducted for additional interfaces to explore other prospective material combinations. This may inform the choice of interface systems and the design of future Majorana experiments. More broadly, similar DFT simulations of interfaces may be performed to evaluate prospective tunnel barriers \textit{e.g.,} for semiconductor devices.

%%%%%%%%%%%%%%%%%%%%%%%%%%%%%%%%%%%%%%%%%%%%%%%%%%%%%%%%%%%%%%%%%%%%%%%%%%%%%%%%%%%%%

\section{Acknowledgements} 
We thank Guang Bian from the University of Missouri, Li Fu from Northwestern Polytechnical University, China, and Tai C. Chiang from the  University of Illinois at Urbana-Champaign for sharing their ARPES data for CdTe. Work at the University of Pittsburgh was supported by the Department of Energy through grant DE-SC-0019274. Work at CMU and UCSB was funded by the National Science Foundation (NSF) through grant OISE-1743717. Work in Grenoble is supported by the ANR-NSF PIRE:HYBRID, Transatlantic Research Partnership and IRP-CNRS HYNATOQ. This research used computing resources of the University of Pittsburgh Center for Research Computing, which is supported by NIH award number S10OD028483 and of the National Energy Research Scientific Computing Center (NERSC), a U.S. Department of Energy Office of Science User Facility operated under Contract No. DE-AC02-05CH11231.

    %\bibliographystyle{unsrt}
    % \bibliography{DFTBibFile}
    \bibliography{etdbib.bib}

\end{document}